\documentclass[a4paper,twoside]{article}
\newcommand{\affil}[1]{$^{\rm #1}$}
\usepackage[authoryear]{natbib}
\bibpunct{(}{)}{;}{a}{}{,}
\usepackage{graphicx}
\usepackage{url}
\date{} 
\title{\large\bf\flushleft Astrophysical Supercomputing with GPUs: 
Critical Decisions for Early Adopters\thanks{Research undertaken as part of the 
Commonwealth Cosmology Initiative (CCI: www.thecci.org), an international collaboration supported by the Australian Research Council}} 
\author{\parbox{\textwidth}{\flushleft
\vspace{-0.5cm}
{\it 
Christopher J.\ Fluke\affil{A,B}, David G. Barnes\affil{A}, 
Benjamin R.\ Barsdell\affil{A} and Amr H.\ Hassan\affil{A}}\\
\vspace{0.4cm}
{\small \affil{A}\,Centre for Astrophysics and Supercomputing, Swinburne University of Technology, PO Box 218, Hawthorn, Australia, 3122}\\
{\small \affil{B}\,Email: cfluke@swin.edu.au}}}
\begin{document}
\twocolumn[
\begin{changemargin}{.8cm}{.5cm}
\begin{minipage}{.9\textwidth}
\vspace{-1cm}
\maketitle
%
%
\small{\bf Abstract:}
General purpose computing on graphics processing units (GPGPU) is 
dramatically changing the landscape of high performance computing in
astronomy.  In this paper, we identify and investigate several
key decision areas, with a goal of simplyfing the early adoption of 
GPGPU in astronomy.  We consider the merits of OpenCL as an open standard 
in order to reduce risks associated with coding in a native, vendor-specific 
programming environment, and present a GPU programming philosophy based on 
using brute force solutions.
We assert that effective use of new GPU-based supercomputing facilities
will require a change in approach 
from astronomers. This will likely include improved programming training, 
an increased need for software development best-practice through 
the use of profiling and related optimisation tools, and a greater 
realiance on third-party code libraries. 
As with any new technology, those willing to take the risks, and
make the investment of time and effort to become early adopters of
GPGPU in astronomy, stand 
to reap great benefits. 


\medskip
\medskip
\end{minipage}
\end{changemargin}
]
\small

\section{Introduction}
\label{sec:intro}

Over the last few decades, astrophysical computation has benefited 
greatly from Moore's Law increases in processing speed -- in 
essence, a doubling of central processing unit (CPU) speed every two years.  
Once a code has been implemented, astronomers have been able to access faster 
processing power by taking advantage of improved hardware as it becomes 
available, but with minimal additional code development. Unfortunately, as 
single-core CPU speeds have plateaued (see Barsdell et al. 2010, 
Figure 1), this scientific software development `free lunch' is 
coming to an end.   However, a radical change in computing architecture 
is providing orders of magnitude improvements in performance, 
and opportunities exist for astronomers to benefit.  

The graphics processing unit (GPU) has appeared as 
a viable, low-cost alternative to traditional CPU computation.  
Indeed, most modern computers now contain a GPU, either as part of 
the main system board or as a peripheral graphics card, and graphics 
hardware performance is doubling on 6-9 month timescales.
In simplest terms, a GPU is a low-cost, highly-parallel coprocessor 
with a high memory bandwidth, which supports single
and double-precision floating point calculations via an instruction set.
Whereas much of the physical chip area of a CPU is devoted to 
control logic and low latency cache memory, GPUs maximise 
the number of processing units that can be accommodated on a chip.

The gradual change in computer graphics architectures from a 
high-cost, fixed function rendering pipeline (early 1980s) for 
graphics-only tasks, to configurable, and ultimately programmable, 
low-cost formats (i.e. GPUs), led to a recognition that 
non-graphics computation on these devices 
was possible (Fournier \& Fussell 1988; Kirk \& Hwu 2010).  
The notion of general-purpose computing on 
GPUs (GPGPU) has now changed the landscape of scientific computation 
across a broad range of disciplines (Tomov et al 2003; 
Venkatsubramanian 2003; Owens et al. 2005).

Timely access to high performance computing (HPC) infrastructure
is a critical ingredient for astronomy to progress,
particular in the fields of numerical computation and signal processing.  
Astronomers have been quick to capitalise on the new 
hardware-accelerated approach to computation (e.g. Nyland et al. 2004; 
Schaaf \& Overeem 2004; 
Portegies Zwart et al. 2007; 
Elsen et al. 2007; 
Hamada \& Iitaka 2007; 
Schive et al. 2007; 
Wayth et al. 2007; 
Belleman et al. 2008; 
Ford 2008; 
Harris et al. 2008; 
Moore et al. 2008;
Nyland et al. 2008; 
Szalay et al. 2008; 
Aubert et al. 2009;
Ord et al. 2009; 
Aubert \& Teyssier 2010; Thompson et al. 2010) 
with most authors reporting speed-ups
of $O$(10) -- $O$(100) times the single-core alternatives. 
Indeed, these early successes in astronomical GPGPU have motivated 
major investments in hybrid CPU+GPU supercomputing infrastructure, 
including the Kolob cluster at the University of 
Heidelberg,\footnote{{\tt http://kolob.ziti.uni-heidelberg.de/}} 
the 170 teraflop/s\footnote{1 flop = 1 floating point operation; 
1 flop/s = 1 floating point operation/second.} Silk Road 
facility\footnote{{\tt http://silkroad.bao.ac.cn/}} operated 
by the NAOC, and the planned Australian GPU Supercomputer for Theoeretical 
Astrophysics Research (gSTAR) with a design goal of $\sim600$ teraflop/s.

The dramatic processing speed-ups that can be achieved by moving 
computation from the CPU to a GPU do not come without costs.  
Early use of GPUs for scientific computation required code to be written 
in graphics card-native shader languages [e.g. Portegies-Zwart et al. (2007)
used Cg for an N-body implementation] such that an arbitrary
computation needed to be recast as if it was a graphics computation relating
to shading pixels and polygon vertices. While the advent of the 
CUDA programming library for NVIDIA
hardware, and the OpenCL standard from the Khronos Group has somewhat 
simplified the task of writing GPU-specific code,  in general, adoption
of GPUs requires a fundamental change in algorithm design and 
implementation. Most critically, the move from straightforward, single-core CPU 
sequential programming to complex, many-core massively parallel stream 
processing may call for radical redevelopment of software, rather 
than simply porting code to a new architecture and recompiling 
(e.g. Owens et al. 2005; Che et al. 2008; Christadler \& Weinberg 2010; 
Larus \& Gannon 2010).  
A detailed analysis 
of HPC systems and applications used by members of the European PRACE 
Consortium (Simpson, Bull \& Hill 2008), which included astrophysical codes,
highlighted the need to rewrite key algorithms and kernels to scale software
effectively to the many-thousand processing cores of petaflop/s systems;
the additional work required to optimse codes for hardware accelerators; and
the need for additional personnel to undertake the coding effort.
While these findings were not directly addressing GPU computing in astronomy, 
they are representative of the challenges that face astronomers who hope 
to take advantage of the massively-parallel processing paradigm. 

HPC is currently navigating a `multi-core corner', marking a 
transition between past (single-core) and future (many-core) architectures.
Indeed, new multi-core technologies are appearing at a rapid rate:
the trend is to see products from competing vendors leapfrog each other 
in processing speed and features supported (e.g. more processing cores, 
greater memory bandwidth, error-correcting memory, 
simultaneous execution of multiple kernels) in the rush to 
maximise market uptake through ever lower `price-per-performance'.  
While the GPU market is dominated by two vendors, NVIDA and AMD, 
these are not the only options for parallel coprocessors. Both 
the Cell architecture,
used in Sony's PlayStation 3 game console, and  field-programmable gate
arrays (FPGA) offer similar order-of-magnitude speed-ups compared to CPU
for certain classes of problems.  Notwithstanding its use as the processor of 
choice for the world's first petaflop/s supercomputer, 
Roadrunner\footnote{{\tt http://www.lanl.gov/}}, 
the future of Cell is uncertain.  In astronomy,  FPGAs are 
better-suited to digital signal processing and radio astronomy applications, 
rather than general purpose computation.  We do not discuss either of these
multi-core options further, but focus our attention on GPUs.

In this paper, we consider two key questions that early adopters of 
GPGPU in astronomy will need to address: choice of programming language
(section \ref{sct:code}), as this impacts on the astronomer's ability to write
any code for GPU, and selection of starting point for implementation
(section \ref{sct:implement}), which may benefit from a return to `brute
force' solutions.  In section \ref{sct:other}, we comment on some 
additional factors that must also be considered by early adopters: 
numerical precision; code optimisation and profiling; and opportunities to
use third-party GPU software libraries.  We present our concluding 
remarks in section \ref{sct:conclusion}.

The choice of suitable problems for implementation on GPUs is beyond the
scope of this work, and is addressed in detail in Barsdell 
et al. (2010). For the remainder of this paper, we assume that the 
reader's code/algorithm of interest has already been identified as 
suitable for a GPU.  Additionally,
it is not our intention to describe all of the features of GPU architectures
or  discuss general programming techniques for massively parallel processors.
Instead, we refer the interested reader to 
Owens et al. (2005)\footnote{Published 
prior to the release of CUDA, some of the implementation issues they 
raise have been resolved}, Che et al. (2008) and Kirk \& Hwu (2010). For 
CUDA code developers, there are resources such as the GPU Gems 
series\footnote{Online versions of volumes 1--3 are freely available 
from {\tt http://developer.nvidia.com/page/home.html}}.

Successful utilisation of the GPGPU paradigm in astronomy relies 
on one main ingredient: software.  The somewhat sobering reality is that
existing CPU-only codes will not run on GPUs without either 
adaptation, re-writing, or a greater reliance on third-party software 
libraries.  A critical question that early adopters of GPGPU will need to 
consider is how to best utilise their limited resources (e.g. time, personnel) 
in order to have science-ready GPU codes.

\section{Software development kits for GPU programming}
\label{sct:code}

A compiler and driver library is required for developing and using GPU
program code. The compiler is a standard C or C++ compiler supporting
a small set of language extensions that are used to declare and define
functions ({\em kernels}) that execute on the GPU, while the driver
library provides standard C or C++ functions for launching or
executing kernels and managing the memory on GPUs.  

\subsection{CUDA and OpenCL}

The NVIDIA Compute Unified Device Architecture (CUDA)\footnote{NVIDIA
  CUDA: \url{http://www.nvidia.com/object/cuda_home_new.html}} is the
prevailing Software Development Kit (SDK) that provides a compiler and
driver library for GPUs.\footnote{Other architecture-specific SDKs
  include the ATI Stream SDK (AMD) for programming ATI Radeon GPUs and
  the Cell Broadband Engine SDK (IBM).} GPU programs written using
this SDK are customarily referred to as CUDA programs; hereafter, we
use CUDA to refer interchangeably to the language extensions and to
the SDK itself. CUDA was released in June 2007 and with very few
exceptions \citep[eg.][]{pzwart07} has been the enabling
technology for the direct application of GPUs in astronomy. 
CUDA has become the de facto standard SDK for astronomy computation as
it is relatively mature and robust, has a large user
community, and comes with extensive documentation and sample
code. CUDA works with all modern NVIDIA GPUs and has been used to
derive worthwhile speed-ups in codes throughout astronomy
as reviewed in Section~\ref{sec:intro} of this paper.

CUDA is not without shortcomings. CUDA programs execute only on NVIDIA GPU
hardware: using the CUDA SDK as a development platform forces one to
choose NVIDIA hardware for execution. Not only are other GPU vendor
solutions (eg. AMD/ATI graphics cards) incompatible with CUDA, other
generic coprocessor hardware (eg. Sony/IBM Cell BE processors) and
standard multi-core CPU processors also lack CUDA
support. Importantly, the CUDA language extensions are not an
openly-defined standard: NVIDIA can change the CUDA capabilities and
interface without notice and at their sole discretion.

The Open Compute Language (OpenCL)\footnote{Khronos OpenCL:
  \url{http://www.khronos.org/opencl/}} is a new, open standard that 
addresses these two deficiencies.  OpenCL defines a hardware-agnostic
application programming interface (API) for general-purpose computing
on GPU hardware.  The OpenCL standard is developed and maintained by
the Khronos Group\footnote{Khronos: \url{http://www.khronos.org/}}
(who also publish the OpenGL standard), but implementations of OpenCL
are provided by third-parties, typically hardware vendors. At the time
of writing, several major OpenCL implementations are available,
variously supporting the four contemporary monolithic processor
architectures:
\begin{itemize}
\item NVIDIA OpenCL supports NVIDIA GPUs and x86 CPUs.
\item AMD OpenCL supports AMD (ATI) GPUs and x86 CPUs.
\item Apple OpenCL supports AMD (ATI) and NVIDIA GPUs and x86 CPUs.
\item IBM OpenCL supports the POWER, Cell Broadband Engine, and x86
  processors.
\end{itemize}
It is necessary to choose a particular OpenCL implementation to build
and execute OpenCL codes, but in contrast to CUDA code, standard-compliant
OpenCL code should compile and execute with {\em any} OpenCL implementation,
and is therefore hardware- and vendor-agnostic.

The OpenCL driver interface and kernel language extensions are very
similar to those of CUDA.  Indeed, the changes required to convert a
simple CUDA code to an OpenCL code are usually limited to the
initialisation and memory copy operations, and minor syntax
differences in the kernel(s).  Instances where more substantial work
might be needed are highly-optimised CUDA kernels, or kernels
requiring or using intricate memory operations for performance or
problem-size related reasons.

Considering that OpenCL is an open standard developed by an industry
consortium, and that several OpenCL implementations are already
available from the mainstream processor vendors, we contend that
OpenCL is more future-proof than CUDA, and is an attractive choice for
GPU development.

\subsection{Performance}
Being more general than CUDA, OpenCL cannot express features specific
or unique to a particular processor. 
Moreover, there {\em will} be GPU kernels that run faster when written 
in a hardware-native API, particularly if the code 
undergoes extensive optimisation
for a particular GPU family. However, for practical use of GPUs in astronomy,
we favour generality and longevity of code over the last, say, ten per
cent performance gain. With this in mind, we now examine the relative
performance of OpenCL and CUDA.

References to CUDA in the astronomy research literature outnumber
those to OpenCL. On 2010 June 17, the SAO/NASA Astrophysics Data
System Abstract Service (Astronomy and Astrophysics Search option) 
returned 51 articles with the text `CUDA' in
the abstract, but only four articles with the text `OpenCL'. Of
these, two report on explicit performance comparison between
CUDA and OpenCL on the same hardware. For the calculation of
gravitational wave source models, \citet{khanna10} measure identical
performance for CUDA and OpenCL on an NVIDIA Tesla GPU. 
Karimi et al. (2010) 
compare implementations of a Monte Carlo simulation for a quantum
spin system, using an NVIDIA GeForce GTX-260 GPU.  Examining 
the relative difference in run-times (their Figure 5) shows 
that performance does depend on the problem size, with variations up to 70\%,
but decreasing to 10-20\% as the complexity of the simulation
increases.  SiSoftware, a
UK-based company providing benchmarking software, have also reported on OpenCL
performance compared to CUDA.\footnote{SiSoftware CUDA and OpenCL
  comparison:
  \url{http://www.sisoftware.info/?d=qa&f=gpu_opencl&l=en&a=}} They
find arithmetic and memory performance to be within ~5 per cent on for
CUDA compared to OpenCL on NVIDIA hardware. 

To address the lack of published OpenCL performance measures within
physics and astronomy, we undertook our own {\em simple}\/ comparison
of CUDA and OpenCL performance for a basic N-body kernel.  Our N-body
GPU kernel implements the standard, direct-force calculation for an
N-particle system evolving in its own gravitational potential; force
integration is applied on the CPU by the classic FORTRAN driver
program {\tt NBODY1} \citep{aarseth99}.  Numerous GPU kernels that
implement the N-body direct force calculation exist, including sample
codes shipped with the NVIDIA CUDA and AMD OpenCL SDKs.  For this work
we wrote our own kernel, which sacrifices {\em some}\/ speed for
algorithm clarity.

To use a GPU kernel from {\tt NBODY1} we wrote a wrapper function in
the C language to replace the inline FORTRAN force calculation code.
Three versions of the wrapper function were produced: a CPU version
using a manual conversion of the FORTRAN code to C; a GPU version
using our own CUDA kernel and driver code; and a GPU version using our
own OpenCL kernel and driver code.  The CUDA and OpenCL kernels and
driver code are algorithmically and functionally identical;
differences in the code are only present due to syntactic requirements
of the kernel compilers, and the different initialisation, memory
management and kernel execution functions as provided by the APIs.

A summary of the direct force N-body kernel is as follows: device
(GPU) memory for particle positions (3-vectors), particle forces
(3-vectors) and particle masses (scalars) is allocated once only, on
the first call to the force calculation wrapper function.  On every
call, current values of the particle positions and masses are copied
to the device memory, and the force calculation kernel is invoked.  We
divide the work up so that the total force on a single particle $F_i$
is calculated by the single thread with thread index $i$; within this
thread the component forces (the forces from individual particles) are
calculated sequentially for a block of particles $\mathcal{B}$ at a
time, whose positions and masses are copied to the fast, shared memory
on the GPU.  The same block of particles $\mathcal{B}$ is used by
other threads ($i+1, i+2, \ldots$) running concurrently to calculate
the (partial) integrated forces on other single particles $F_{i+1},
F_{i+2}, \ldots$.

We executed our GPU-enabled {\tt NBODY1} code over a range of particle
number ($N$) sufficient to show scaling behaviour, and on three GPU
systems:
\begin{itemize}
\item MAC8800: an Apple Mac Pro workstation and NVIDIA 8800GT graphics card,
  using NVIDIA-supplied CUDA version 2.3, and Apple-supplied OpenCL
  [version included with Snow Leopard 10.6.2];
\item SMTESLA: a Supermicro Linux workstation with NVIDIA GT200 (Tesla
  C1060) card,
  using NVIDIA-supplied CUDA (with OpenCL) version 2.3; and
\item PCRADEON: a PC clone workstation with AMD ATI Radeon HD 5970
  graphics card,\footnote{Only one of the two GPUs on the Radeon card
    were used in these tests.} using AMD-supplied ATI Stream SDK 2.01
  with OpenCL.
\end{itemize}

\begin{figure*}
\includegraphics[width=6in]{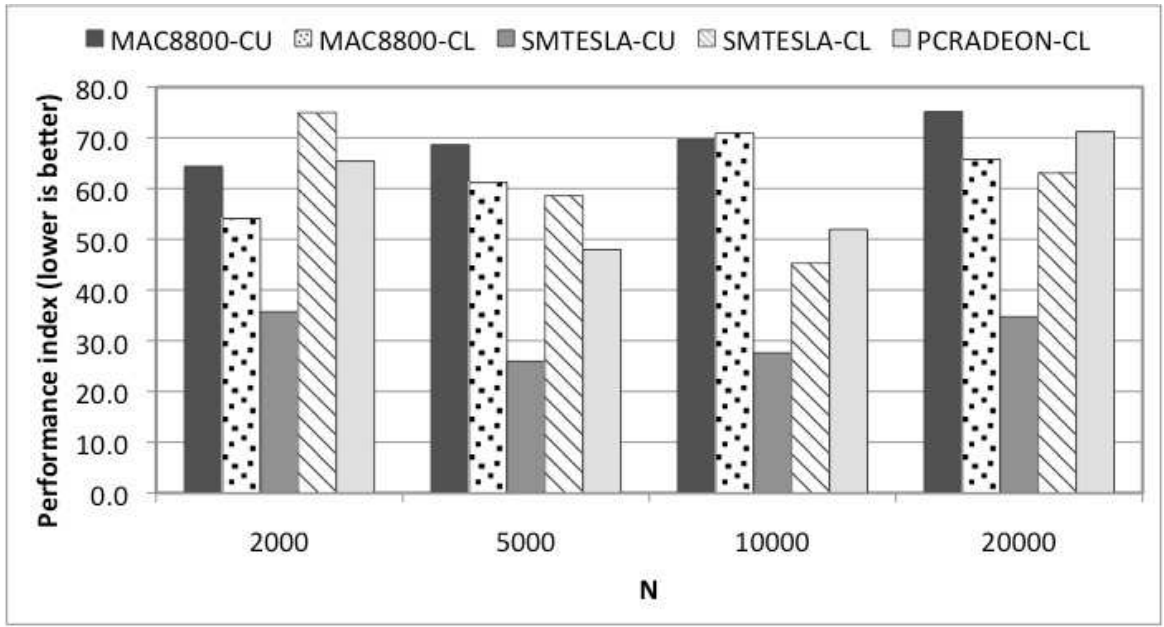}
\begin{centering}
  \caption{Performance index [measured execution time scaled by
    $(N/1000)^{3/2}$] for the {\tt NBODY1} code using CUDA
    (MAC8800-CU, SMTESLA-CU) and OpenCL (MAC8800-CL, SMTESLA-CL,
    PCRADEON-CL) kernels to directly calculate the interparticle
    forces, for varying system sizes.  Lower performance index is
    better.}
\label{fig:perf}
\end{centering}
\end{figure*}

\mbox{Figure \ref{fig:perf}} presents our results.  We note that the 
domain of $N$ values explored here does not fully utilise the capabilities
of the GPUs.  The performance index is the real-world time taken to 
reach the same simulation time (\mbox{50~Myr}), scaled by $(N/1000)^{3/2}$. 
We are illustrating performance comparisons rather than
how performance scales with $N$.  
On the MAC8800 system, the OpenCL kernel executed at least as quickly,
and typically slightly quicker by up to 15 per cent, than the CUDA
kernel.  On the SMTESLA system, the CUDA kernel was faster than the
OpenCL kernel by between 40 and 55 per cent.  These results are not
surprising: Apple Inc.\ markets OpenCL as a major performance feature
of the Snow Leopard operating system, and has evidently delivered a
mature OpenCL implementation that is competitive with CUDA.  On the
Linux-based SMTESLA system, the OpenCL implementation is significantly
younger than CUDA; the MAC8800 results make it reasonable to expect
the NVIDIA OpenCL implementation to mature and become competitive with
CUDA on the Linux platform.  In any case, the performance difference
between CUDA and OpenCL on the SMTESLA system is relatively minor when
one considers that the OpenCL implementation on the SMTESLA system
already out-performs an OpenMP implementation of the kernel executing
on an Intel Xeon E5345 2.33~GHz processor with four processing cores,
by 50 times.

Results for the PCRADEON system running {\tt NBODY1} using an OpenCL
kernel are also shown in \mbox{Figure \ref{fig:perf}}: its performance
is within 10 per cent of that of the OpenCL kernel executing on the
SMTESLA system.  Considering that the ATI HD 5970 card retails for
less than half the price of the Tesla C1060 card, and we have only
used one of its two GPUs in this test, this is an outstanding result.
Using both GPUs on the ATI HD 5970 can be expected to deliver
performance at the level of the SMTESLA system using a CUDA kernel,
for jobs that fit in the smaller memory ($< 2$ GB) of the ATI card.
There is clearly now a choice of hardware {\em and}\/ software vendor
for scientific computing using GPUs.  Choosing OpenCL does not bestow
a significant\footnote{i.e.\ the difference between OpenCL and CUDA
  kernel execution performance is typcially a factor or 10--100
  smaller than the the gain achieved by using GPUs instead of CPUs.}
performance degradation compared to CUDA, and opens up additional
options in GPU hardware for astronomy.

\section{A GPU Programming Philosophy}
\label{sct:implement}
Having considered the merits of CUDA and OpenCL for developing GPU-enabled 
astronomical software, we now consider a related question: for a given
computational problem that has been identified as suitable for a GPU {\em 
which version of the algorithm should be written or adapted for GPU?}  
We present here a GPU-programming philopsophy based on brute force 
implementations.

\subsection{Benefits of Brute Force}
As a starting point for our GPU programming philosophy, we consider
results from Thompson et al. (2010) and Bate et al. (2010) for the
specific case of gravitational microlensing ray-shooting. 
Microlensing is the study 
of the gravitational deflection of light by compact lens masses within 
more massive, extended, lens galaxies; ray-shooting is a computational
technique that follows light rays backwards from an observer, through a 
system of lenses that deflect the light rays, and onto a 
grid in the source plane.  This grid, or magnification map, is then used
to make statistical comparisons with microlensing observations. 

The `industry standard' ray-shooting method for single-core CPU 
uses a sophisticated tree-code (Wambsganss 1990; 1999).  A tree hierachy 
is used to approximate the mass distribution as a collection of pseudo-lenses,
in order to avoid the computational requirements of a direct summation
over all lenses.  This approach was originally 
introduced to microlensing two decades ago due to the unfeasibly 
long processing times (i.e. months to years) that the conceptually simpler, 
brute force computation using all lenses, would take
on the best single-core CPUs available at the time.  However, ray-shooting
is inherently parallel -- each light ray can be deflected independently of
all other light rays, and the deflection due to each microlens adds
linearly -- thus making it an ideal candidate for a GPU implementation.
Today, a high-resolution brute force calculation 
can be achieved in a matter of hours on a GPU. Moreover, with an NVIDIA 
Tesla S1070 unit, Thompson et al. (2010) achieved billion-particle 
microlensing calculations at over 1 teraflop/s and runtimes of a few days.  

Bate et al. (2010) compared the single-core tree code ({\tt CPU-T}) with 
the brute force approach on GPU ({\tt GPU-D} where {\tt D} = direct) 
over a range of 
astrophysically-motivated parameter space. Overall, runtimes using 
{\tt GPU-D} were found to be {\em no worse} than for {\tt CPU-T}: indeed, for 
certain combinations of parameters, {\tt GPU-D} was a factor of a few faster. 
The implementation time for {\tt GPU-D} using CUDA was quite short -- 
a matter of weeks -- compared to the anticipated  time of several months
to implement a working, optimised tree code.\footnote{We note that at the time
of code development, OpenCL had not been publically released, hence our choice of CUDA.}

What we infer from this case is that a na\"ive, simple to implement, 
and more accurate brute force solution is highly competive with a 
clever, complex, fast, trusted, industry standard code.  Since its first 
release, the {\tt CPU-T} code has not required significant modification in order 
to achieve faster processing times. Instead, it has been able to rely almost
solely on the Moore's Law increase in processing speed (and a corresponding 
growth in CPU memory at reduced cost). However, the plateau in CPU clock 
rates means that no additional performance improvement can occur for {\tt CPU-T}
in its present form.  {\tt GPU-D} is now ready to take advantage of the 
anticipated increases in GPU speeds in the years ahead, with
no additional code development required.  The massively parallel 
GPU architecture means that a brute force approach that was not feasible 
even a decade ago is now highly competitive with the algorithmically-complex 
approach.  However, the latter was the only approach that was feasible 
for single-core CPU computing.

The Bate et al. (2010) result suggets an intriguing approach 
to GPU programming that we encourage other early adopters to consider.  
In order to adapt code to GPU, the two main alternatives are:
\begin{enumerate}
\item Take existing code and port it {\em as is} to GPU.  This is likely to 
be a time-consuming task, as many aspects of complex implementations do not
match well with the hardware architecture and memory management requirements of
current GPUs; or
\item Think about what the code is currently doing and why [e.g. 
using an algorithm analysis approach as in Barsdell et al. (2010)].
\end{enumerate}	
If the latter reveals that a complex algorithm was being used in order
to overcome a pre-existing hardware (i.e. single-core CPU) limitation, 
consider the advantages of a brute force approach. This may entail looking back
to how a problem was originally posed, and taking the simple solution --
provided it exhibits the required massive parallelism in computation that
is suited for GPU.

Significant processing speed-ups can be obtained by going beyond a na\"ive 
implementation, but at the same time, a simple algorithm may yield unexpected
speed-ups by over-computing. A case in point is the pairwise force 
calculation in  direct $N$-body gravity simulations: saturation 
of GPU threads can be achieved by introducing additional particles 
with zero mass, and by explicitly calculating the pairwise 
forces $F_{ij}$ and $F_{ji}$, even though $F_{ij} = F_{ji}$ 
(Belleman et al. 2008).  Aubert \& Teyssier (2010) describe their use of
an explicit time integration method for solving the equations of radiative
transfer, taking advantage of GPU acceleration to remove a limitation 
that this approach would otherwise introduce.

Based on experiences so far, simples codes that are more accurate, more
intuitive, and easier to implement for GPU, may result in runtimes that are 
already no worse than the best currently available (i.e. single-core)
codes.  We propose that brute force techniques where the programmer
does not need to be as aware of computer science techniques, using
algorithms that will hopefully map to a more obvious implementation on GPU,
and which can be achieved over a short period of time (see section 
\ref{sct:time}), are a sensible starting point for early adoption.  
In the longer term, once programmers have mastered the details of stream
processing, and if the speed-up offered by a brute force solution is
still not sufficient or scales poorly with the problem size, 
more sophisticated solutions can of course be implemented.

\subsection{Brute-force multi-dimensional minimisation}

To further test this philosophy, we consider a common task in
astronomy: finding the global minimum of a multi-dimensional dataset.
Standard techniques, such as steepest descent, simulated annealing or
the simplex method (described below), attempt to limit the number of
function evaluations required to obtain the minimum.  While
convergence to a solution can occur rapidly, there is no certainty
that the global minimum has been obtained, rather than a local
minimum. Moreover, solutions are often strongly dependent on the
starting point for evaluation.  Whereas techniques exist to find
starting values that bracket the location of a minimum for
one-dimensional functions, no such bracketing techniques exist for the
general multi-dimensional case.

\subsubsection{Downhill simplex minimisation}

A popular multi-dimensional minimisation 
technique is the downhill simplex method (DSM),
introduced by Nelder \& Mead (1965).  A practical implementation of 
this algorithm is is provided with the GNU Scientific library 
(GSL\footnote{{\tt http://www.gnu.org/software/gsl}}) as the function
{\tt gsl\_multimin\_fminimizer\_nsimplex}. DSM is a very general
multi-dimensional minimisation algorithm, as it does not depend on 
knowledge of the derivatives of a function (such as is required for 
the steepest descent algorithm), and hence is appropriate for a wider 
range of applications.   DSM works as follows: based on an initial guess 
at a solution, $\mathbf{p}_0$, an additional $N$ vectors are generated 
using a stepping vector, $\mathbf{s}$.  An $N$ dimensional simplex 
is constructed from these $N+1$ vectors as vertices, and the function
evaluated at each vertex. A set of geometrical transformations are applied 
to the simplex in an attempt to span the minimum value,  at which point 
the simplex contracts in size.  This process is continued iteratively
until a stopping criterion is reached.

\begin{figure}
\includegraphics[scale=0.20, angle=0]{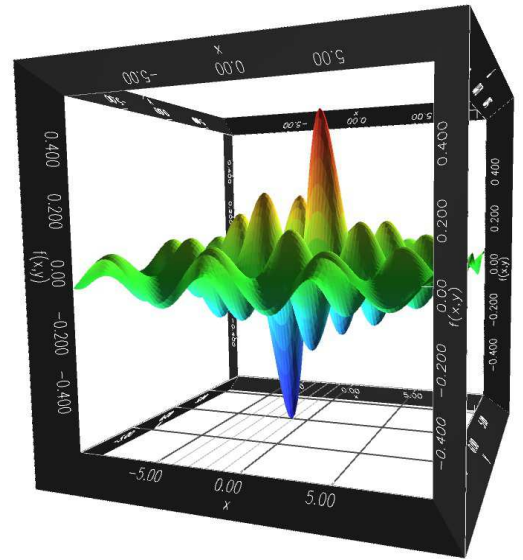}
\caption{Surface representing the function 
$f(x,y) = \left[\sin(x) \cos(y)\right] /\sqrt{1.0+x^2 + y^2}$. A rainbow
colour-range (blue to red) is used to highlight function values. 
Three-dimensional visualisation is perhaps the easiest way to identify
global minima and maxima for (well-behaved) two-dimensional functions! }
\label{fig:toymodel}
\end{figure}

As a demonstration of the difficulties associated with using DSM, 
consider the following well-behaved function in two-dimensions:
\begin{equation}
f(x,y) = \frac{\sin(x)\cos(y)}{\sqrt{1.0+x^2+y^2}}
\end{equation}
which is plotted in Figure \ref{fig:toymodel}.  As the visualisation 
ably demonstrates, there is one unique global minimum 
in the range $x \in [-10,10], y \in [-10,10]$, at $(x,y) = (-1.11, 0.0)$ 
and a number of local minima.

\begin{figure*}
\begin{center}
\includegraphics[scale=0.32, angle=0]{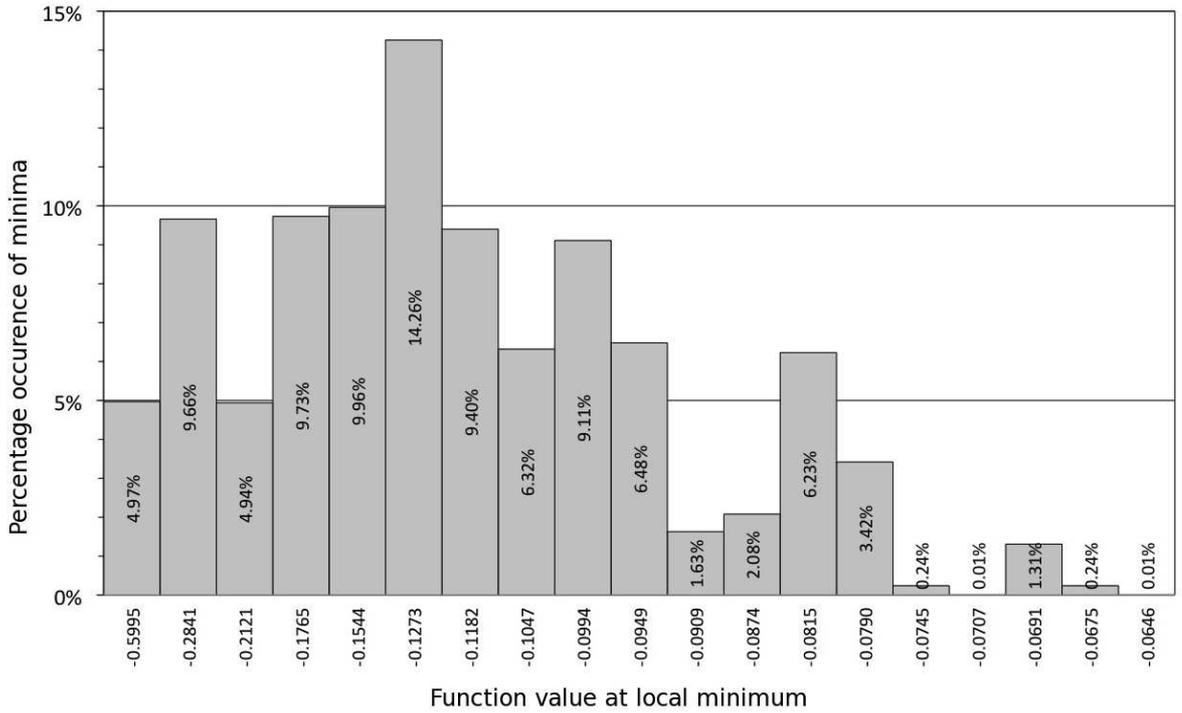}
\end{center}
\vspace{-0.6cm}
\caption{Distribution of minima found by GSL simplex routine for minimisation
of equation (1), with starting values in the range $[-10,10]$ for $(x,y)$. 
Results based on executing code $10^4$ times. The actual 
global minimum is only found in $\sim 5\%$ of runs.}
\label{fig:distribution}
\end{figure*}

We implemented a wrapper for the DSM function in C, and 
ran $10^4$ iterations with random starting values for $(x,y)$ over 
the range $[-10,10]$.  
Total runtime for $10^4$ DSM solutions on a Xeon 5138 
processor\footnote{A single processor of a 2x quad-core Clovertown 
Processor.} at 2.33 GHz was 0.4 s.  
Figure \ref{fig:distribution}
shows the distribution of minima returned by the simplex routine.  
While DSM does indeed find the global minimum, it only achieves
this in $5\%$ of cases.\footnote{We note that a na\"ive use of the 
Mathematica {\tt Minimize} function also returns an incorrect global minimum
value $f(1.4454,28.89434) = -0.284$.}  

Next, we trial a $5$-dimensional function
\begin{equation}
\label{eq:fn}
g(x,y,z,v,w) = \frac{\sin(x)\cos(y)\sin(z)\cos(v)\sin(w)}
{\sqrt{1 + x^2 + y^2 + z^2 + v^2 + w^2}},
\end{equation}
which cannot be visualised as easily as for equation (1).  However,
by noting the symmetries $x \leftrightarrow z \leftrightarrow w$ and
$y \leftrightarrow v$, we can consider the function:
\begin{equation}
h(u) = \frac{\sin^3(u)}{\sqrt{1 + 3u^2}},
\end{equation}
which has a minimum at $u = -1.36672$, and deduce that there are four global
minima at $(y,v) = (0,0)$ and $(x,z,w)$ as listed in Table \ref{tbl:minfun}. 
\begin{table}
\caption{Global minima of the five-dimensional function of equation (2) 
occuring when $y=v=0$.}
\label{tbl:minfun}
\begin{tabular}{rrrr}
$x$ & $z$ & $w$ & $f(x,y,z,v,w,)$  \\ \hline
-1.36672 & 1.36672 & 1.36672 & -0.365412\\
1.36672 & -1.36672 & 1.36672 & -0.365412\\
1.36672 & 1.36672 & -1.36672 & -0.365412\\
-1.36672 & -1.36672 & -1.36672 & -0.365412 \\
\end{tabular}
\end{table}
As before, the DSM code was executed $10^4$ times.  Average runtime 
(over 10 runs) on the same CPU architecture as used previously
was 1.97 s.  For the five-dimensional function, one of the four global minima 
was only found in 0.03 $\%$ of cases.

As a simple modification, we enable the code to find successively lower
values.  Starting with an initial guess, DSM converges to a minimum value.
A new starting point is chosen, and the solution is compared to the previous
best minimum.  This process continues until no lower minimum has been found
after 100 iterations.  This change increased the runtime to 
5.23 seconds (average over 10 runs), and resulted in one of the four 
global minima being found in $0.13 \%$ of runs. 

An alternative strategy is to reduce the range for starting guesses,
although in general, we might not know {\em a priori} what an appropriate 
(reduced) range is. Here, by choosing starting values in the range 
$[-5,5]$, and using the additional iteration step, our code returns
a correct global minimum in $3.7 \%$ of runs. A secondary effect here
is a slight reduction in runtimes (4.8 s).  

We have now demonstrated the difficulties with using an approximate
method, such as DSM, for minimimisation of a well-behaved function.
Our intent is not to provide a detailed analysis of how to improve
performance of DSM, but to present a starting point for considering an
alternative minimisation technique suitable for GPU: brute-force
computation over a search domain.

\subsubsection{Brute-force minimisation}

In brute-force method (hereafter, BFM) minimisation, function values
are evaluated on a grid of points in multi-dimensional parameter
space. The resulting array of function values is searched or sorted to
identify the minimum (or minima). The principal characteristic of this
approach is that the quality of the solution---how close the result is
to the {\em real}\/ minimum (or minima)---depends on the resolution of
the grid.  In general, a finer grid yields a grid point (or set of
grid points) closer to the true global minimum (minima). However, 
for real-world minimisation, minima will rarely align exactly 
with a grid point and consequently, unlike DSM, the BFM on its own 
may not yield the {\em exact}\/ minimum.

What BFM {\em can}\/ do though is identify the location, with known uncertainty
of one-half of the grid spacing in each direction, of {\em physically
  meaningful}\/ minima.  By this we mean minima that are smooth with
respect to the selected grid resolution, or put another way, minima that are
well-behaved over the neighbouring grid cells.  A sharp negative
spike, confined in the multi-dimensional parameter space to a
hypervolume smaller than a grid cell, cannot reliably be identified by
brute-force; i.e.\ high-frequency features may be missed by the sampling of
the parameter space.

Happily, for many real-world minimisation problems, especially those
using measured data, it is actually quite reasonable and
straightforward to set a physically meaningful resolution for
searching parameter space.  For example, consider fitting a Gaussian
profile to a radio continuum image of a galaxy, with the position,
width and amplitude parameters of the Gaussian to be determined.  In
most cases, position will be over-sampled at one-tenth of
the resolution (beam size) of the image (although for extremely bright
sources needing the best astrometry this could be refined); width
would be sufficiently sampled in multiples of one-fifth the beamsize;
and amplitude in multiples of one-fifth of the image noise.  While
this sampling implies a desired precision in the final fit, if
subdividing the grid much finer than this yields substantially
different solution(s), it is likely to be due to unphysical artefacts
in the image (e.g. sharp spikes from interference or noise outliers, which 
are not convolved with the observing beam) rather than genuine
signals of interest.

Minima found by brute-force sampling of parameter space can be used in
their own right, or used as initial guesses for a method like DSM
which could refine the solution (i.e.\ move it off the grid and
towards the {\em exact}\/ minimum).  BFM minimisation can easily
accommodate {\em a priori}\/ conditions on the parameter search space,
such as invalid regions (which can be masked out from the search) or
parameters requiring varying resolution (e.g.\ using log sampling).

To explore the feasiblity of BFM minimisation using GPUs, we
implemented the evaluation of equation (\ref{eq:fn})
over a defined parameter space using an OpenCL kernel and driver
program.  For simplicity, all five variables were sampled over the
domain $(-5,5)$ sampled at $N$ points; a total of $N^5$ function
evaluations are required.  For a given triplet $(x_i,y_j,z_k)$, the
kernel was programmed to sequentially evaluate the function over all
$(v,w)$, a total of $N^2$ evaluations, 
and return the vector $(x_i,y_j,z_k,v_{\rm
  min},w_{\rm min})$ and value $g_{\rm min}(x_i,y_j,z_k)$ where the
minimum function value was identified for the triplet $(x_i,y_j,z_k)$.
The kernel was deployed over a 3-dimensional work block $\mathcal{B}$
of size $(N,N,N)$, itself divided into work groups of size $(N,1,1)$.
The driver program received $N^3$ values of $g_{\rm min}(x,y,z)$ at
the completion of the kernel invocation, and used sequential CPU code
to identify the global minimum and its location.

\begin{figure*}
\includegraphics[width=6in]{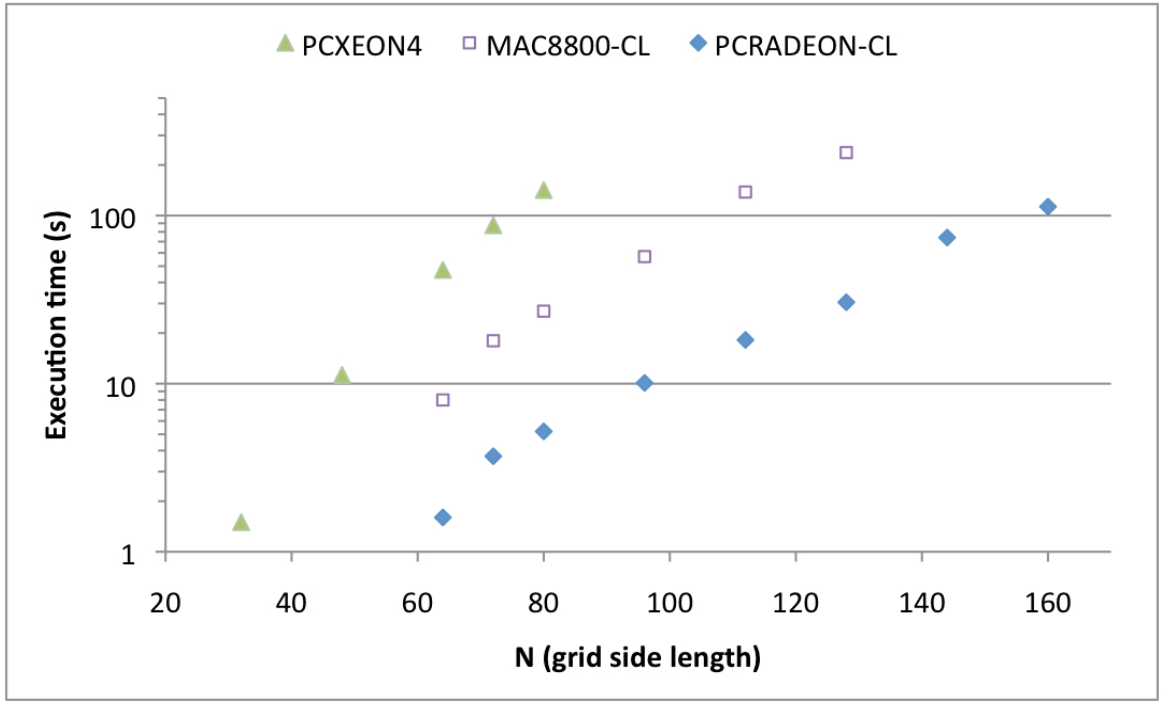}
\begin{centering}
  \caption{Measured execution time for the brute-force evaluation of
equation (\ref{eq:fn}) on a $N^5$-cell grid for a quad-core Xeon
processor (PCXEON4) and OpenCL (MAC8800-CL, PCRADEON-CL) kernels.
Lower execution time is better.}
\label{fig:brute}
\end{centering}
\end{figure*}

We measured the time taken to evaluate the test function for $N$ in the
range 32--160.  We used the MAC-8800 and PCRADEON systems as
previously described, as well as a quad-core Intel Nehalem i7 930 Xeon
system (PCXEON4) running a standalone C implementation (five nested
loops over the $x,y,z,v,w$ parameter axes) of the brute force
calculation.\footnote{The standalone C implementation was a
  single-core code; we report quad-core timings by assuming perfect
  scaling which is reasonable for this task.}  \mbox{Figure
  \ref{fig:brute}} presents the results.  For this compute-bound
problem, the PCRADEON system (using 1 GPU) outperforms the PCXEON4
system by around 25 times, and can evaluate the function ($g$) at a
rate of $\sim10^{9}$ evaluations per second.  Using both GPUs on the
PCRADEON system would double this speed.

As previously noted, equation (\ref{eq:fn}) has multiple minima,
yet it is highly improbable that the grid used by the BFM
exactly aligns with any of them.  However, provided the grid is
sampled finely enough, BFM minimisation will identify a point
close to one of the four global minima, as the minimal point on the
grid, hereafter $P_b$.  As the sampling of the grid increases (i.e. as
$N$ increases) the {\em overall}\/ expectation is that $P_b$, and the
function evaluated at that point, $g(P_b)$, will approach the position
and value of (one of) the global minima respectively.  However this
approach is not necessarily piecewise continuous because the minimum
function evaluation on the grid will likely flip-flop from
side-to-side of local or global minimum as $N$ is increased.

\begin{figure*}
\includegraphics[width=6in]{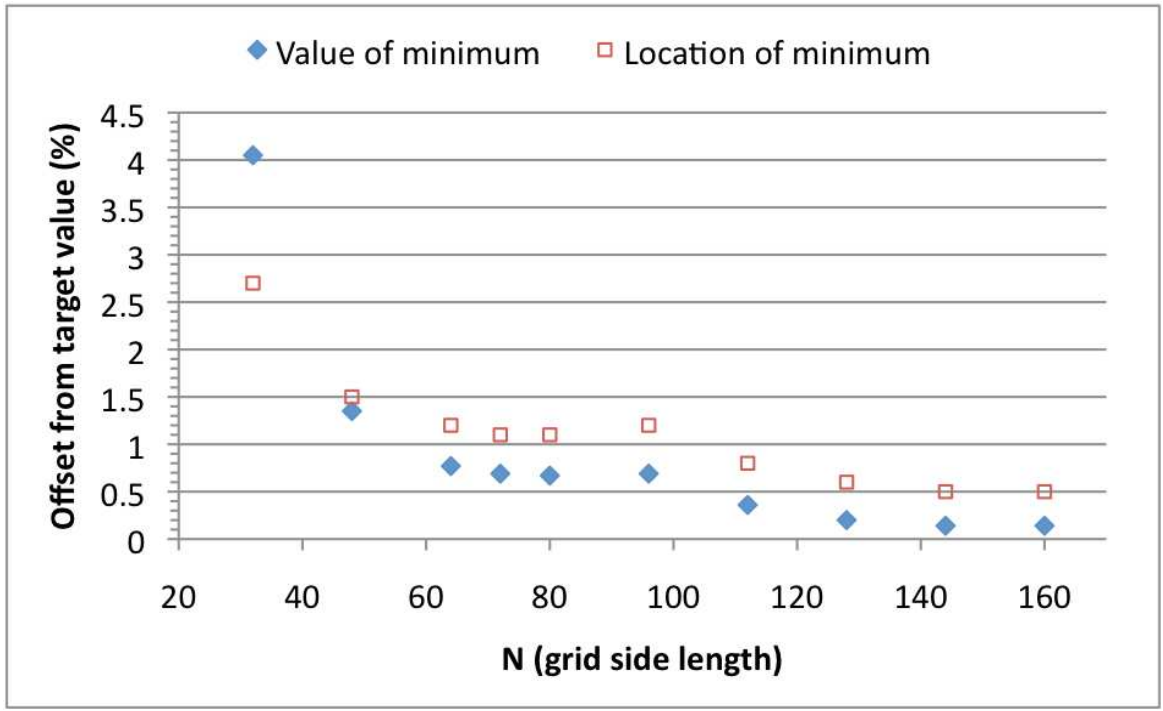}
\begin{centering}
  \caption{Convergence of BFM minimisation to the value and
    location of a global minimum of equation (\ref{eq:fn}) as a
    function of the grid side length $N$.  The value offset is
    expressed as a percentage of the target minimum value; the
    position offset as a percentage of the diagonal length of the
    5-dimensional parameter space.}
\label{fig:goodness}
\end{centering}
\end{figure*}

Figure~\ref{fig:goodness} illustrates the convergence
towards an acceptable solution for the test function with increasing
subdivision of parameter space---for $N \simeq 100$, the global minimum
value is accurate to $\sim 0.5$~per~cent and the position is within
$\sim 1$~per~cent of one of the known global minima positions.  In the real
world, with e.g.\ measurement noise and resolution effects added to
this dataset, $N$ would be sensibly limited by known properties of the 
data.

To compare the standard downhill simplex method to the brute force
approach for minimising the test function, we consider what can
be achieved by each method in \mbox{10 s}:
\begin{itemize}
\item DSM (on the Xeon 5138 CPU) has a $7.7$~per~cent strike rate in
  identifying {\em exactly} one of the four global minima (i.e.\ it
  has a $92.3$~per~cent likelihood of {\em not}\/ identifying one of
  the global minima!); and
\item BFM (on the PCRADEON-CL system) identifies a lowest value within
  $0.7$~per~cent of the known global minimum value, lying closer than
  $1.2$~per~cent of the diagonal length of the parameter space to one
  of the known global minima locations.
\end{itemize}
While BFM minimisation could be re-run at higher resolution in a
sub-volume of parameter space around minima found on lower resolution
grid/s to improve the accuracy of its results, a hybrid DSM-BFM method
is obviously suggested.  Using for example the $\sim100$ lowest points
found in the brute force evaluation of the function $g$ over the
5-dimensional parameter space with $N=64$, as initial guesses for the
downhill simplex method, we obtain a method which can readily identify
all four global minima for the function $g$, exactly, in just a few
seconds.  

Or intent in this section has been to demonstrate that a brute force
approach is a practical {\em starting point} for a GPU, which gives 
a significant speed-up compared to a CPU (Figure \ref{fig:brute}), 
and may indeed
overcome some of the existing limitations with the optimised alternatives.  
Implementating BFM minimisation using CUDA or OpenCL is straightforward, 
whereas efficient coding of a more complex minimisation algorithm 
for a GPU is likely to be a time-consuming task -- and would result 
in the same issues with identifying local rather than global minima. 
The brute force approach could indeed be used in a multi-scale 
fashion to obtain higher accuracy, e.g. for each local minima, 
repeat the brute force evaluation but on a higher resolution local grid, 
or for providing the starting points for an alternative techinque.  

\subsection{Time to Science}
\label{sct:time}
Since there is an overhead in preparing code
to run effectively and optimally on GPU, it is worth considering how much time
is spent on programming versus the actual speed-ups.  
For the astronomer, whose primary interest is their ability to advance 
knowledge through computation, rather than advancing knowledge of the
computational technique, we can consider a time to science, 
$T_{\rm science}$, to aid decisions regarding adoption of GPU:
\begin{equation}
T_{\rm science}= T_{\rm learn} + T_{\rm implement} + T_{\rm run} + 
T_{\rm analysis}.
\end{equation}
Here, $T_{\rm learn}$ is the time taken to learn the fundamentals of
a programming approach (CPU or GPU), $T_{\rm implement}$ is the time
to implement a specific programmatic solution, $T_{\rm run}$ is 
the runtime of
the code, and $T_{\rm analysis}$ is the time taken to analyse the outputs
in preparation for future work or publication.  A GPGPU approach is
a desirable or beneficial in cases where $T_{\rm science}(GPU) 
\ll T_{\rm science}(CPU)$, but can we estimate how each factor 
is likely to contribute?

The most apparent advantage of GPGPU, and the most significant factor
in $T_{\rm science}$, is the expectation that
\begin{equation}
T_{\rm run}(GPU) \ll T_{\rm run}(CPU).  
\end{equation}
This has been demonstrated for a growing range of GPU-astronomy applications, 
with $10-100\times$ speed-ups in processing time.  If this condition is
not met, there is little to be gained from a GPU implementation. 

For a given number of outputs, we expect that
\begin{equation}
T_{\rm analysis} (GPU) \sim T_{\rm analysis} (CPU),
\end{equation}
as the analysis time should not be affected by
the computational runtime.  If we consider  a scenario where faster GPU 
runtimes are used primarly to produce more outputs than are possible with
CPU, then the total analysis time will grow.

Based on our experiences, we suggest that 
\begin{equation}
T_{\rm learn} (GPU) > T_{\rm learn} (CPU), 
\end{equation}
and may indeed be 
\begin{equation}
T_{\rm learn} (GPU) \gg T_{\rm learn} (CPU). 
\end{equation}
In general, the average astronomer-programmer does not have 
training in parallel programming techniques - we anticipate that this situation
may be resolved in the years ahead as multi- and many-core programming 
makes it way into undergraduate courses, but this may be a major factor 
in the short term.  
While it has been possible for many code-writing 
astronomers to remain unaware of the hardware configuration of a CPU (beyond
knowing the total available memory), optimising code for GPUs requires 
a more complete knowledge of issues such as available memory bandwidth, 
and appropriate allocation of data between register, device and 
shared memory spaces (e.g. Che et al. 2008; Christadler \& Weinberg 2010).

Regardless of architecture, we assert that:
\begin{equation}
T_{\rm implement}({\rm simple}) \ll T_{\rm implement} 
({\rm complex})
\end{equation}
namely that is faster and easier to implement a simple, brute force solution,
than to code a more complex algorithm.  As our initial investigations
suggest, such an approach can still result in:
\begin{equation}
T_{\rm run}(\mbox{GPU-simple code}) \leq T_{\rm run} (\mbox{CPU-complex code}).
\end{equation}

Assessing the combined effects of these factors is challenging, 
however, we suggest that $T_{\rm science}$ will be smaller for simple
to implement, brute-force GPU codes in cases, despite the overheads in 
learning to appropriate GPU programming techniques.  Of course, 
as with traditional CPU programming, these techniques only need to be 
learnt once, and can then be applied to a range of computational problems
in the future. 
In the same way that single-core CPU codes have benefitted 
from Moore's Law, once a simple GPU code is available, it can also 
take advantage of the anticipated processing speed-ups that will 
occur in future generations of GPU hardware, 
regardless of whether additional time/effort is invested to develop
a more optimised `complex on GPU' solution.

\section{Other considerations}
\label{sct:other}
We now briefly comment on several additional factors that early adopters
should be aware of: issues of performance and precision,
tools to aid in profiling and optimising code, and the role 
of third-party GPU-code libraries and GPU-enhanced programming environments.
 
\subsection{Performance, Precision and Optimisation}
Graphics processing generally only requires single precision floating 
point calculations, but astronomy computation may require double 
precision computation in order to achieve sufficient numerical accuracy.  
Presently, there is a big performance difference 
between single-precision (SP) and double-precision (DP) computation, as 
GPU hardware has fewer DP processors.  In some cases, DP emulation can 
be achieved by packing a DP value (64 bit) into two SP floats (32 bit), 
resulting in two times increase in precision -- this approach was used 
by Gaburov et al. (2009) in their {\tt SAPPORO} N-body code.  This is 
likely to only be a short-term limitation for GPUs, as increased DP 
performance is under consideration by GPU vendors (e.g. while 
SP speeds have increased overall, in the move from the NVIDIA Tesla S1070 to 
the newer Fermi hardware, DP speeds have improved from 1/8 to 1/2 the SP rates).
An additional factor that may limit longer calculations is the 
availability of error correcting memory (ECC), which is able to mitigate
failures due to memory errors from interference (including cosmic rays)
or hardware problems [see Schroeder et al. (2009) for an empirical study].
NVIDIA's Fermi cards now support EEC, but with a slight
decrease in processing performance when this mode is enabled, and a
reduction in the amount of allocatable GPU memory.

Writing working GPU code is not the same as writing optimised, efficient
GPU code.  A simple GPU implementation of 
a parallel algorithm, such as those presented here, may require several 
iterations in order to reach a sufficiently optimised version. Factors 
of $2-10\times$ in speed can be achieved quickly, but reaching 
$100\times$ does require effort and expertise. Tools such as the 
CUDA occupancy calculator\footnote{{\tt http://developer.nvidia.com}} 
can help to improve speed through improved use of GPU memory spaces 
and ensuring that all stream processors are being highly utilised.
We caution that quoted peak speeds for GPU hardware are primarily 
for graphics-like calculations (e.g. dual issuing of a 
multiplication and addition per clock cycle for single-precision).  
A more realistic outcome is to achieve about 1/4 to 1/2 of the quoted peak 
performance, at best, but it is strongly problem-dependent.

While many astronomer-programmers may be familiar with techniques for 
debugging code, they may be less aware of the existence or importance of 
code profiling.  A software profiler examines run-time characteristics
of codes, including memory allocation, time spent executing functions,
and the frequency of their execution.  Profiling can help identify
code sections that may benefit from optimisation.  CUDA and 
both the NVIDIA and AMD OpenCL implementations provide visual profilers 
as part of their SDKs.  Investing time learning
to use these tools effectively may have a beneficial impact on 
$T_{\rm science}(GPU)$. 

\subsection{Third-party libraries and programming environments}
\label{sct:thirdparty}
From our own experience, there is a general trend for astronomers to 
re-implement code or algorithms that may already exist in third-party 
libraries.  Obvious exceptions to this include the use of the 
FFTW\footnote{{\tt http://www.fftw.org}}
Fourier transform libraries, the PGPLOT graphics subroutine library 
library\footnote{{\tt http://www.astro.caltech.edu/$\sim$tjp/pgplot}}, and 
code fragments or implementations 
from Numerical Recipes 
Software.\footnote{{\tt http://www.numerical-recipes.com/}} A growing
number of GPU-oriented libraries are now available, including the
CUDA Data parallel Primitives Library 
(CUDPP\footnote{{\tt http://code.google.com/p/cudpp/}}) that 
provides primitives for common tasks like sorting and building 
data structures, and CUFFT -- NVIDIA's own CUDA FFT library.  
 
Alternatives to programming code in CUDA or OpenCL, optimised for execution
on a GPU, include GPU-enabled enhancements to widely used 
interactive environments and scripting languages such as 
the Interactive Data Language (IDL)\footnote{{\tt http://www.ittvis.com/ProductServices/IDL.aspx}}, 
Mathematica\footnote{{\tt http://www.wolfram.com/products/mathematica}}, 
and the Python programming language.\footnote{{\tt http://www.python.org/}} For example,
the CUDA-basd GPULib by Tech-X 
Corporation\footnote{{\tt http://www.txcorp.com/products/GPULib/}} 
allows IDL scripts to access a GPU for common mathematical functions 
and processes (e.g. interpolation, correlation and parallel geneneration 
of random numbers). Similarly, 
PyCUDA\footnote{{\tt http://mathema.tician.de/software/pycuda}} and PyGPU\footnote{{\tt http://www.cs.lth.se/home/Calle\_Lejdfors/pygpu}} are two implementations that enable GPU computing within Python.  Moreover, compiler-based
solutions such as the HMPP Workbench\footnote{\url{http://www.caps-entreprise.com}}, aim to hide the details of
GPU code development through the use of OpenMP-like directives in standard codes. As with the 
case of chosing OpenCL in preference to CUDA, there is likely to be some processing 
overhead in using one of these solutions, but they do provide a simpler, 
high-level access to GPU that may be preferable for many types of 
applications.

\section{Concluding Remarks}
\label{sct:conclusion}
In this paper, we have highlighted some of the benefits 
and limitations of early adoption of GPGPU for astronomy.  While there 
are risks and significant effort may be required to prepare codes, in many
cases the benefits will outweigh the limitations.
A preferred outcome for astronomers 
is a majority of time and effort spent on scientific outcomes rather 
than software development.

The promise of the OpenCL standard, 
is to provide opportunities for hardware-agnostic coding.  
OpenCL seems to present a good amount 
of flexibility for implementation, rather than using a native API 
(such as CUDA for NVIDIA), without a significant decrease in processing speed.  
Furthermore, we suggest that
for certain classes of scientific computations a step backwards 
to consider simple, brute force solutions that were not feasible for 
CPU, may instead reduce software development times. The resulting codes 
may already be `no worse' than the best single-core alternatives, and 
may even be more accurate, or overcome limitations of existing 
optimised approaches. Lessons learnt in starting with brute force solutions
can then help researchers to determine whether a longer-term solution does
indeed warrant the effort of implementing a more sophisticated alternative.

While running codes faster may be an end in itself, faster computation means 
that there is more time to explore parameter space.  This might include 
running models with different parameters, or running repeat models with 
different random seeds in order to build up a more robust
statistical sample. Additionally, GPUs provide opportunities to tackle 
computational problems that are still not feasible on single core CPU 
or traditional multi-core computing clusters, at a greatly reduced
cost.  

Not all applications require GPUs, so some time and effort should 
be invested in understanding the types of problems that will really 
achieve the greatest benefit.  For example, telescope control
software does not parallelise well, if at all, but the highly 
parallel nature of Fourier transformation, used extensively in astronomy, 
makes it an ideal candidate for GPGPU.  Indeed, there are problems,
such as the conceptually simple process of generating a histogram from
data values, that are easy to implement on a CPU, but which 
become unnecessarily complex when attempting to find a parallel solution.
Computational tasks compatible with 
a stream processing paradigm, i.e. many individual data-streams 
requiring identical computations, are candidates for moving from the CPU 
to the GPU.   Fortunately, a high degree of data parallelism is
present in many astronomy scenarios (e.g. the use of the CLEAN algorithm 
in radio astronomy, which takes advantage of data parallelism in the
spectral domain).

Identification of {\em relevant} astronomy computations is 
the first step towards implementation on GPU. Barsdell et al. (2010)
propose an approach based on algorithm analysis, whereby common 
processing tasks are matched to a taxonomy of algorithms. Despite
its obvious application to GPGPU development, this approach 
may provide insight into improvements and optimisation for single-core
and multi-core CPU codes.  An improved understanding of the
parallelism in existing astronomy codes and algorithms can lead to simple
optimisation, such as identifying sections of code that would benefit
from the trivial parallelisation on multi-core architectures possible with 
OpenMP.  Indeed, in some cases, the simpler shared-memory programming 
model provided by OpenMP may provide a sufficient processing improvement
without resorting to a GPU solution.  For a code to be moved
to a GPU, rather than using OpenMP on a modern quad-core architecture, 
it typically needs to present a 20-30x speed-up over the 
single-core solution. The counter argument is that GPUs remain 
significantly cheaper on a \$/gigaflop basis than a sufficiently 
multi-core computer (e.g. $> 24$ cores) that would provide
the same processing performance. 

We emphasise that skills in parallel or stream processing programming 
techinques are not widespread amongst astronomy graduates or graduate 
students, and the `supervisor teaching the student to code' may no 
longer be feasible.  It is often a difficult decision to choose between spending
limited research resources on direct (travel, graduate student stipends, 
postdoctoral salaries) versus indirect research costs (programmers who may
not have, nor actually seek, scientific training).  However, it will not 
be feasible for astronomers to prepare for GPU-based HPC facilities 
without some investment in training on parallel programming techniques. 

The long-term role of the GPU is still unknown: whether they will
remain as a computational coprocessor, or if multi-core CPUs will grow 
to become more GPU-like.  There may be other radical changes in hardware 
in the years ahead, such as the experimental 48-core Intel single-chip 
cloud computer announced in 2009.  The recent demise of Intel's Larabee 
consumer GPU chip, a hybrid CPU and GPU, with features such as 
cache coherency across all cores and greater flexibility in computation, 
may have delayed resolution of this issue for at least a few more years.  
While these
short term changes may lead to some redundancy in code development effort, 
awareness of the fundamental differences between CPU and GPU programming 
and execution should provide insight into problem solving for 
future highly parallel architectures.
Moreover, we anticipate that the move to astronomical GPGPU may not 
be limited to HPC facilities, but will ultimately encompass desktop and 
notebook supercomputing.

GPGPU represents a natural new direction for astrophysical HPC.  Adoption
of a radical new processing architecture, and the correspoding 
required change in approach to software development, is worthwhile 
if our understanding of the universe advances at an accelerated rate. 
We remain enthusiastic about the prospects for GPGPU in astronomy.

\section*{Acknowledgments} 
This research was supported under Australian Research Council's 
Discovery Projects funding scheme (project number DP0665574).
We are grateful to Jarrod Hurley and Matthew Bailes for discussions 
regarding GPGPU, and to our referee for insightful comments regarding
this work.

\end{document}